\documentclass[10pt,twocolumn]{ieeeconf}
\usepackage{amssymb}
\usepackage{amsmath}
\usepackage[version=3]{mhchem}
\usepackage{graphicx}
\usepackage{multirow}
\usepackage{wrapfig}
\usepackage{stackengine}
\usepackage{color}
\usepackage{colortbl}
\usepackage{soul}
\usepackage{subfigure}
\usepackage{enumerate}
\usepackage{paralist}
\usepackage[normalem]{ulem}
\usepackage{mathtools}

\begin{document}
\title{\textbf{On the Capacity of Diffusion-Based Molecular Communications with SiNW FET-Based Receiver}}
\author{Murat Kuscu,~\IEEEmembership{Student Member,~IEEE}
        and Ozgur B. Akan,~\IEEEmembership{Fellow,~IEEE}
        \thanks{The authors are with the Next-generation and Wireless Communications Laboratory (NWCL), Department of Electrical and Electronics Engineering, Koc University, Istanbul, 34450, Turkey (e-mail: \{mkuscu, akan\}@ku.edu.tr).}
        \thanks{This work is a part of the Project MINERVA supported by the European Research Council (ERC) under grant ERC-2013-CoG \#616922.}}% <-this % stops a space
\maketitle
\thispagestyle{empty}
\pagestyle{empty}

% NOTES: each receptor binds to only a single ligand. monovalent? no cooperativity between them. independent observations.

\begin{abstract}
Molecular communication (MC) is a bio-inspired communication method based on the exchange of molecules for information transfer among nanoscale devices. Although MC has been extensively studied from various aspects, limitations imposed by the physical design of transceiving units have been largely neglected in the literature. Recently, we have proposed a nanobioelectronic MC receiver architecture based on the nanoscale field effect transistor-based biosensor (bioFET) technology, providing noninvasive and sensitive molecular detection at nanoscale while producing electrical signals at the output. In this paper, we derive analytical closed-form expressions for the capacity and capacity-achieving input distribution for a memoryless MC channel with a silicon nanowire (SiNW) FET-based MC receiver. The resulting expressions could be used to optimize the information flow in MC systems equipped with nanobioelectronic receivers.
\end{abstract}

\IEEEpeerreviewmaketitle

\section{Introduction}
\label{Introduction}
Molecular communication (MC) is a promising means of realizing nanonetworks and the Internet of nano-bio things (IoBNT), which could find breakthrough applications in several domains, such as continuous health monitoring and smart drug delivery \cite{Akan2016}, \cite{Kuscu2015}. This bio-inspired method is mainly based on the use of molecules to encode, transmit and receive information.

MC has been studied from several perspectives, including that of information theory \cite{Akyildiz2013}, \cite{Nakano2012}. However, except for a few studies, the design of transceiving units and corresponding physical limitations have been largely neglected in the literature. There exist mainly two approaches to design MC transceivers. The first approach is to use synthetic biology techniques for designing artificial cells capable of transmitting and receiving molecular messages \cite{Unluturk2015}. The second one is based on nanobioelectronic architectures which exploit the extraordinary properties of novel nanomaterials, such as nanowires and graphene, with a bio-inspired approach. In our recent review of design options for the MC receiver, we justify that nanobioelectronic designs are feasible and more convenient in most cases \cite{Kuscu2016a}. Electrical output of a nanobioelectronic receiver provides the advantage of high-speed information processing, making the design more suitable for the complex communication schemes and protocols developed for MC \cite{Nakano2012}.
% mostly based on biosensors

\begin{figure}[!t]
\centering
\includegraphics[width=7cm]{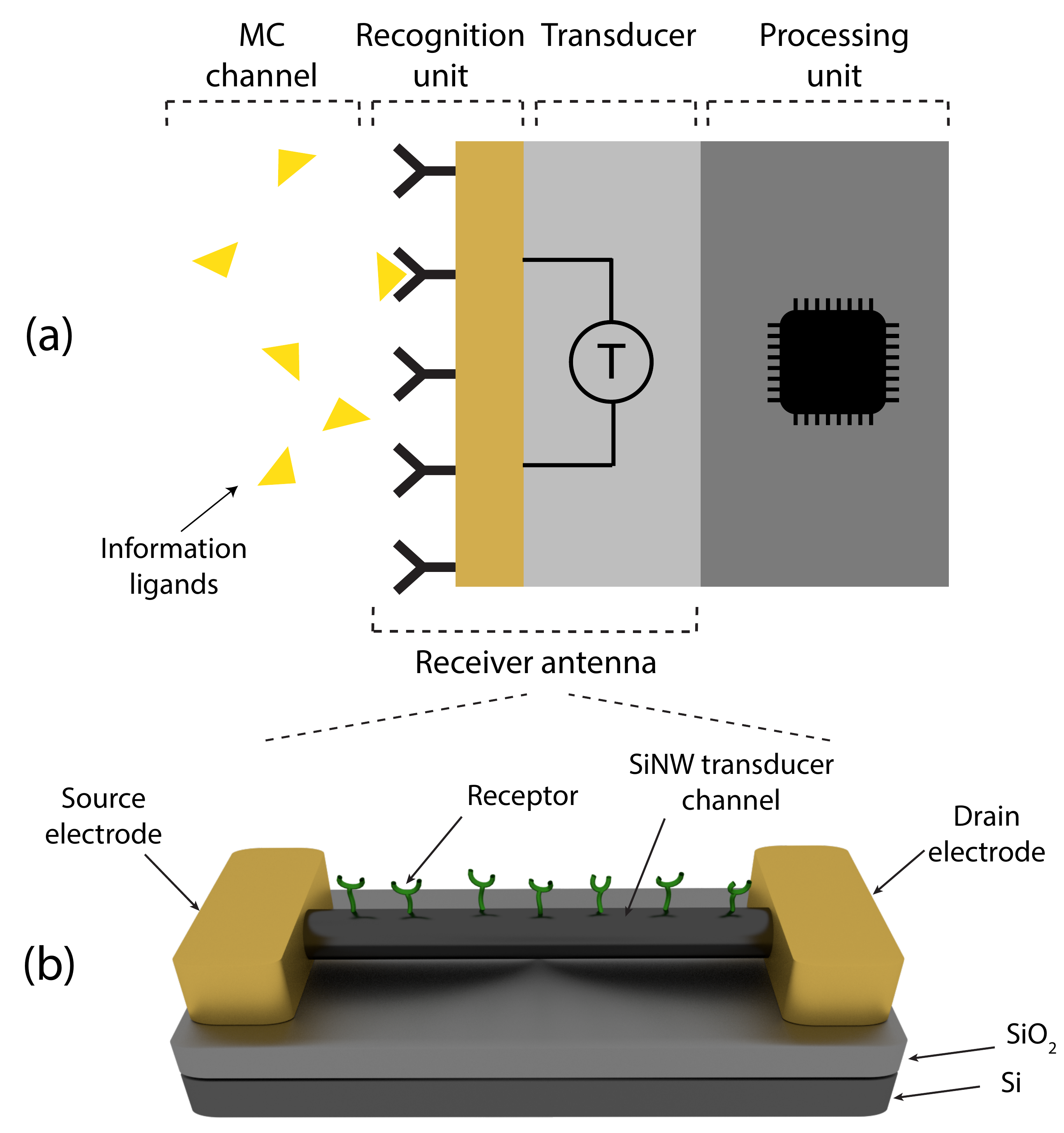}
\caption{(a) Functional units of an MC receiver, and (b) SiNW FET-based MC receiver antenna.}
\label{fig:molrec}
\end{figure}

In \cite{Kuscu2016b}, we develop a communication theoretical model for a SiNW FET-based MC receiver and obtain the noise statistics. This nanobioelectronic design, shown in Fig. \ref{fig:molrec}, is based on SiNW FET biosensors (bioFETs), which operate in a similar way to the conventional FETs. A SiNW bioFET consists of a biorecognition layer of surface receptors capable of binding the target ligands and a SiNW transducer channel which transduces the intrinsic charges of bound ligands into an electrical current flowing between the source and drain electrodes. In this paper, we develop an information theoretical model for a memoryless MC system equipped with the proposed SiNW FET-based MC receiver. We derive analytical closed-form expressions for the MC capacity and capacity-achieving input distribution, following the methodology of \cite{Marzen2013}. The resulting expressions incorporate the effect of transmitter, channel and receiver parameters, and could be used for optimizing the information flow in MC systems including nanobioelectronic receivers.

The remainder of the paper is organized as follows. In Section II, we present the existing model of MC system with a SiNW FET-based receiver. The capacity is derived in Section III, and the results are provided in Section IV. Finally, the concluding remarks are given in Section V.

\section{MC System Model}
\label{model}
We consider a time-slotted MC system between a single transmitter nanomachine (TX) and a single receiver nanomachine (RX). TX and RX are assumed to be perfectly synchronized with each other in terms of time. Information is encoded into the number of molecules, i.e., ligands, released by the TX. The channel is assumed to be memoryless. This can be practically realized with a low symbol rate or using auxiliary enzymes degrading the residual molecules that cause inter-symbol interference (ISI).

The channel is assumed to be a free diffusion channel, where the trajectories of individual ligands are independent of each other. Thus, the ligand concentration as a function of time and space can be expressed by the Fick's equation
\begin{equation}
\frac{\partial \rho(\vec{r}, t)}{\partial t}=D \nabla^2 \rho(\vec{r},t), \label{ficks}
\end{equation}
where $\rho(\vec{r}, t)$ is the ligand concentration at location $\vec{r}$ and time $t$, and $D$ is the diffusion coefficient of the ligands. TX is assumed to be a point source located at the origin of 3D Cartesian coordinate system, i.e., $\vec{r_T} = (0,0,0)$. It releases $N_{tx} $ number of molecules at the very beginning of a time slot. Thus, we can express the initial condition as $\rho(\vec{r}, 0) = N_{tx} \delta(\vec{r})$. Assuming that the distance $d$ between the TX and RX surface is much higher than the largest dimension of the RX surface, we can consider that all of the surface receptors are at equal distance from TX, and thus, are exposed to the same ligand concentration, which can be obtained by solving the Fick's equation at the RX location with the initial condition imposed by the TX:
\begin{equation}
\rho(d, t) = \frac{N_{tx}}{(4 \pi D t)^{3/2}} \exp\left(-\frac{d^2}{4Dt}\right).
\end{equation}
We assume that RX samples the occupation states of the receptors when the ligand concentration at the receiver location is at its maximum, which is the case when $t = \frac{d^2}{6D}$ \cite{llatser2011}. The corresponding maximum ligand concentration at the RX location is then given by
\begin{equation}
\rho_R = \left( \frac{3}{2 \pi e}\right)^{3/2} \frac{N_{tx}}{d^3} = \alpha_{ch} \times N_{tx},
\end{equation}
where we define $\alpha_{ch} = (3/2 \pi e)^{3/2} d^{-3}$ as the channel attenuation parameter.

The interaction of ligands with the surface receptors is governed by the ligand-receptor binding kinetics. Assuming steady-state conditions, the probability of finding a single receptor at the bound state is given by
\begin{equation}
P_b = \frac{k_1 \rho_R}{k_1 \rho_R + k_{-1}} = \frac{k_1 \alpha_{ch} N_{tx}}{k_1 \alpha_{ch} N_{tx} + k_{-1}}, \label{p1slow}
\end{equation}
where $k_{-1}$ is the unbinding rate constant for the receptor-ligand pair; and $k_1$ is the binding rate constant in $m^3 s^{-1}$. Generalizing the results to the case with $N_r$ receptors, and assuming that all of the receptors observe independent concentrations, the distribution of number of bound receptors becomes binomial \cite{Berezhkovskii2013}:
\begin{equation}
P(N_b = k) = \binom{N_r}{k}(P_b)^{k} (1-P_b)^{N_r-k},  \label{condprob}
\end{equation}
with mean and variance given by
\begin{equation}
\mu_{N_b} = P_b N_r, \quad \sigma_{N_b}^2 = P_b (1-P_b) N_r. \label{meanNb}
\end{equation}
%\subsection{Transduction}
Surface potential created on the NW-electrolyte interface as a result of binding of a single ligand is given as
\begin{equation}
\psi_L = (q_{eff} \times N_e^-)/C_{eq},
\end{equation}
where $N_{e^-}$ is the number of free electrons per ligand molecule, and $C_{eq}$ is the equivalent capacitance of the transducer. $q_{eff}$ is the mean effective charge of a free ligand electron, which is degraded as the distance between the ligand electron and the transducer increases due to the Debye screening. The relation is given by $q_{eff} = q \times exp(-r/\lambda_D)$, where $q$ is the elementary charge, and $r$ is the average distance of ligand electrons in the bound state to the transducer's surface \cite{Rajan2013}, which is assumed to be equal to the average surface receptor length, i.e., $r = l_{SR}$. The Debye length, $\lambda_D$, quantizes the ionic strength of the medium according to
\begin{equation}
\lambda_D = \sqrt{\frac{\epsilon_M k_B T}{2 N_A q^2 c_{ion}}}, \label{eq:debye}
\end{equation}
where $\epsilon_M$ is the dielectric permittivity of the medium, $k_B$ is the Boltzmann's constant, $T$ is the temperature, $N_A$ is Avogadro's number, and $c_{ion}$ is the ionic concentration of the medium \cite{Rajan2013}. The transducer capacitance is given by
 \begin{equation}
C_{eq} = \left(\frac{1}{C_{ox}} + \frac{1}{C_{nw}} \right)^{-1} + C_{dl}, \label{eq:debye}
\end{equation}
where $C_{dl}$ is the diffusion layer capacitance resulting from the double layer created by the medium counterions accumulated at the interface between oxide layer, i.e., SiO$_2$ layer, and the electrolyte medium; $C_{ox}$ is the capacitance of the oxide layer; and $C_{nw}$ is the SiNW capacitance, which is again a double layer capacitance caused by the accumulation of carriers to the SiO$_2$/SiNW interface \cite{Gao2010}, \cite{Shoorideh2014}.

For a nanowire-on-insulator (NWoI) configuration of RX, we can consider the SiNW as a hemicylinder with an oxide layer of thickness $t_{ox}$ covering the surface, and a diffusion layer of thickness $\lambda_D$ covering the oxide layer \cite{Shoorideh2014}. Accordingly, the diffusion layer capacitance can be given by $C_{dl} = (\epsilon_{M}/\lambda_D) w_R l_R.$, where $w_R = \pi r_R$ is the effective width of SiNW with radius $r_R$, and $l_R$ is the SiNW length. The oxide capacitance is given by $C_{ox} = C_{ox}^s w_R l_R = (\epsilon_{ox}/t_{ox}) w_R l_R,$ where $\epsilon_{ox}$ is the permittivity and the thickness of the oxide layer, respectively. We will use $C_{ox}^s$ to denote the oxide capacitance per area. For high values of hole density, e.g., $p \sim 10^{18}$ cm$^{-3}$, corresponding to the linear operation regime of the FET, the $C_{nw}$ is obtained as $C_{nw} = (\epsilon_{Si}/\lambda_{nw}) w_R l_R$ \cite{Gao2010} , where $\epsilon_{Si}$ is the dielectric permittivity of SiNW, and $\lambda_{nw}$ is the thickness of the double layer created in the inner surface of the NW, which is given by $\lambda_{nw} = \sqrt{\epsilon_{Si} k_B T/p q^2}$.

The potential induced at the SiNW/oxide layer interface is reflected into the output current of the RX. We assume that the p-type FET is operated in the linear (Ohmic) region; thus, its transconductance is expressed by $g_{FET} = \mu_p C_{ox}^s V_{SD} w_R/l_R$, where $V_{SD} = -V_{DS}$ is the source to drain voltage of the SiNW FET. Given the transconductance, the output current, $I_{rx}$, resulting from the surface potential generated by the bound ligands is given as $I_{rx} = g_{FET} \psi_L N_b$, which has the mean
\begin{equation}
\mu_{I_{rx}} = g_{FET} \times \psi_L \times \mu_{N_b}. \label{Imean}
\end{equation}

The operation of bioFET-based MC receiver is suffered from $1/f$ noise. Based on \emph{the correlated carrier number and mobility fluctuation model}, power spectral density (PSD) of the $1/f$ noise at the output current can be written as
\begin{equation}
S_{I_{rx}^F}(f) = S_{V,FB}(f) g_{FET}^2 \left[1 + \alpha_s \mu_p C_{ox}^s (V_{SG} - |V_{TH}|)  \right]^2 , \label{eq:Is}
\end{equation}
where $V_{SG} = -V_{GS}$ is source to gate voltage, and $V_T$ is the threshold voltage of the SiNW FET. $\alpha_s$ is the Coulomb scattering coefficient, and $\mu_p$ is the mobility of the hole carriers \cite{Rajan2010}. The PSD of the flatband-voltage noise $S_{V,FB}$ is given by
\begin{equation}
S_{V,FB}(f) = \frac{\lambda k_B T q^2 N_{ot} g_{FET}^2}{w_R l_R (C_{ox}^s)^2 |f|},
\end{equation}
where $\lambda$ is the characteristic tunneling distance, $N_{ot}$ is the oxide trap density, i.e., impurity concentration, of the SiNW channel \cite{Rajan2010}. $1/f$ noise is independent of the received signals, and shows an additive behavior on the overall output current fluctuations. Theoretically, $1/f$ noise does not have a low frequency cutoff, and has infinite power at zero frequency. However, in experimental studies with a finite measurement time, a finite variance for $1/f$ noise is observed. The reason is related to the low frequency cutoff set by the observation time $T_{obs}$ \cite{Niemann2013}. Considering that the received molecular signals are at the baseband, to be able to calculate the total noise power, we assume one-year operation time, i.e., $ \sim \pi \times 10^{7}$ s, for the antenna such that the low cutoff frequency is $f_L = 1/T_{obs} \approx  1/\pi \times 10^{-7}$ Hz. At frequencies lower than $f_L$, the noise is assumed to show the white noise behavior, i.e., $S_{I_{rx}^F}(f) = S_{I_{rx}^F}(f_L)$ for $\lvert f \rvert < f_L$.
\begin{equation}
\sigma_F^2 = \int_{-\infty}^{\infty} S_{I_{rx}^F}(f) df.
\end{equation}

We can expect that a significant number of receptors, e.g. $>1000$, are tethered to the top surface of a SiNW channel. For large number of surface receptors, the binomial distribution given in \eqref{condprob} can be approximated as Gaussian, i.e., $N_b \sim \mathcal{N}\left(\mu_{N_b}, \sigma_{N_b}^2\right)$. $1/f$ noise is resulting from the bias current flowing through SiNW channel, thus, it is independent of the binding noise. It can be approximated to follow a Gaussian distribution \cite{Hooge1969}. Therefore, the overall noise process effective on the output current becomes a stationary Gaussian process with the variance
\begin{equation}
\sigma_{I_{rx}}^2 =  \sigma_F^2 + \sigma_{N_b}^2 g_{FET}^2 \psi_L^2 \label{overallvariance}.
\end{equation}

\section{Capacity}
We consider the overall system as a continuous-input continuous-output memoryless channel. Although information is encoded into discrete number of molecules, we can assume that the input is continuous given that it takes values from a very large input range. Then, the mutual information between the transmitted signal $N_{tx}$ and the received signal $I_{rx}$ can be written as

\small
\begin{equation}
\begin{split}
I&(N_{tx}; I_{rx}) \\
&= \int_{N_{tx}^{min}}^{N_{tx}^{max}} f_{N_{tx}}(x) \int_{-\infty}^{\infty} f_{I_{rx}|N_{tx}}(y|x) \log_2f_{I_{rx}|N_{tx}}(y|x) dy dx\\
&- \int_{N_{tx}^{min}}^{N_{tx}^{max}} f_{N_{tx}}(x) \int_{-\infty}^{\infty} f_{I_{rx}|N_{tx}}(y|x) \log_2f_{I_{rx}}(y) dy dx \label{mutual}
\end{split}
\end{equation}
\normalsize
where $N_{tx}^{min}$ and $N_{tx}^{max}$ denote the minimum and maximum number of molecules that TX can transmit. The input probability distribution satisfies the condition $\int_{N_{tx}^{min}}^{N_{tx}^{max}} f_{N_{tx}}(x) dx = 1$. Based on the discussion at the end of Section \ref{model}, overall channel transition pdf $f_{I_{rx}|N_{tx}}$ can be given by
\begin{equation}
f_{I_{rx}|N_{tx}}(y|x) = \frac{1}{\sqrt{2\pi \sigma_{I_{rx}}^2}} e^{-\frac{\left(y-\mu_{I_{rx}}\right)^2}{2\sigma_{I_{rx}}^2}}  \label{condprobgauss}
\end{equation}
where $\mu_{I_{rx}}$ and $\sigma_{I_{rx}}^2$ are functions of $N_{tx} = x$, as given in \eqref{Imean} and \eqref{overallvariance}. Given the Gaussian pdf, the first integral on the RHS of \eqref{mutual} is obtained as \cite{Marzen2013}
\small
\begin{equation}
\int_{-\infty}^{\infty} f_{I_{rx}|N_{tx}}(y|x) \log_2f_{I_{rx}|N_{tx}}(y|x) dy = - \frac{1}{2} \log_2(2 \pi e \sigma_{I_{rx}}^2).
\end{equation}
\normalsize
We approximate the second integral on the RHS of \eqref{mutual} using Taylor series expansion of $\log_2p(\mu_{I_{rx}})$ up to the second order \cite{Marzen2013}:
\small
\begin{equation}
\begin{split}
& \int_{-\infty}^{\infty} f_{I_{rx}|N_{tx}}(y|x) \log_2f_{I_{rx}}(y) dy \\
&= \int_{-\infty}^{\infty} f_{I_{rx}|N_{tx}}(y|x) \log_2f_{\mu}(\mu_{I_{rx}}) dy \\
&+ \int_{-\infty}^{\infty} f_{I_{rx}|N_{tx}}(y|x) \left. {\frac{\partial \log_2f_{I_{rx}}(y)}{ \partial y}}_{\stackunder[1pt]{}{}}%
 \right|_{%
 \stackon[1pt]{$\scriptscriptstyle \mu_{I_{rx}}$}{}}(I_{rx} - \mu_{I_{rx}}) dy \\
&+ \frac{1}{2} \int_{-\infty}^{\infty} f_{I_{rx}|N_{tx}}(y|x)  \left. {\frac{\partial^2 \log_2f_{I_{rx}}(y)}{ \partial y^2}}_{\stackunder[1pt]{}{}}%
 \right|_{%
 \stackon[1pt]{$\scriptscriptstyle \mu_{I_{rx}}$}{}}(I_{rx} - \mu_{I_{rx}})^2 dy. \label{integral}
\end{split}
\end{equation}
\normalsize
The first integral in \eqref{integral} becomes
\small
\begin{equation}
\begin{split}
\int_{-\infty}^{\infty} &f_{I_{rx}|N_{tx}}(y|x) \log_2f_{\mu}(\mu_{I_{rx}}) dy \\
& = \log_2f_{\mu}(\mu_{I_{rx}}) \int_{-\infty}^{\infty} f_{I_{rx}|N_{tx}}(y|x) dy = \log_2f_{\mu}(\mu_{I_{rx}}), \label{integral2}
\end{split}
\end{equation}
\normalsize
and the second integral translates into
\small
\begin{equation}
\begin{split}
& \int_{-\infty}^{\infty} f_{I_{rx}|N_{tx}}(y|x) \left. {\frac{\partial \log_2f_{I_{rx}}(y)}{ \partial y}}_{\stackunder[1pt]{}{}}%
 \right|_{%
 \stackon[1pt]{$\scriptscriptstyle \mu_{I_{rx}}$}{}}(y - \mu_{I_{rx}}) dy  \\
& = \left. {\frac{\partial \log_2f_{I_{rx}}(y)}{ \partial y}}_{\stackunder[1pt]{}{}}%
 \right|_{%
 \stackon[1pt]{$\scriptscriptstyle \mu_{I_{rx}}$}{}} \int_{-\infty}^{\infty} f_{I_{rx}|N_{tx}} (y - \mu_{I_{rx}}) dy = 0. \label{integral3}
\end{split}
\end{equation}
\normalsize
The third integral is simplified as follows
\small
\begin{equation}
\begin{split}
\int_{-\infty}^{\infty} & f_{I_{rx}|N_{tx}}(y|x) \frac{1}{2} \left. {\frac{\partial^2 \log_2f_{I_{rx}}(y)}{ \partial y^2}}_{\stackunder[1pt]{}{}}%
 \right|_{%
 \stackon[1pt]{$\scriptscriptstyle \mu_{I_{rx}}$}{}}(y - \mu_{I_{rx}})^2 dy  \\
& = \frac{1}{2} \left. {\frac{\partial^2 \log_2f_{I_{rx}}(y)}{ \partial y^2}}_{\stackunder[1pt]{}{}}%
 \right|_{%
 \stackon[1pt]{$\scriptscriptstyle \mu_{I_{rx}}$}{}} \int_{-\infty}^{\infty} f_{I_{rx}|N_{tx}}(y|x) (y - \mu_{I_{rx}})^2 dy \\
& = \frac{1}{2} \left. {\frac{\partial^2 \log_2f_{I_{rx}}(y)}{ \partial y^2}}_{\stackunder[1pt]{}{}}%
 \right|_{%
 \stackon[1pt]{$\scriptscriptstyle \mu_{I_{rx}}$}{}} \sigma_{I_{rx}}^2 \approx 0. \label{integral4}
\end{split}
\end{equation}
\normalsize
The approximation in \eqref{integral4} follows from the fact that the variance is very small compared to the mean for the high values of the number of surface receptors \cite{Kuscu2016b}, \cite{Marzen2013}. Using \eqref{integral2}-\eqref{integral4} we can approximate the second integral on the RHS of \eqref{mutual} as follows

\small
\begin{equation}
\int_{-\infty}^{\infty} f_{I_{rx}|N_{tx}}(y|x) \log_2f_{I_{rx}}(y) dy \approx \log_2f_{\mu}(\mu_{I_{rx}}).
\end{equation}
\normalsize
%Mutual information becomes
%\small
%\begin{equation}
%\begin{split}
%I(N_{tx}; I_{rx}) =& - \int_{N_{tx}^{min}}^{N_{tx}^{max}} f_{N_{tx}}(x) \Bigl( \log_2\sqrt{2 \pi e \sigma_{I_{rx}}^2} \\
%&+ \log_2f_{\mu}(\mu_{I_{rx}}) \Bigr) dx
%\end{split}
%\end{equation}
%\normalsize

Since $\mu_{I_{rx}}$ is a deterministic and single valued function of $N_{tx}$, $f_{\mu}(\mu_{I_{rx}})$ can be written in terms of the input distribution $f_{N_{tx}}(x)$ as follows

\small
\begin{equation}
\begin{split}
f_{\mu}(\mu_{I_{rx}}) = f_{N_{tx}}(x) \frac{(x \alpha_{ch} k_1 + k_{-1})^2}{N_r M}, \label{transformedprob}
\end{split}
\end{equation}
\normalsize
where $M = \alpha_{ch} k_1 k_{-1} g_{FET} \psi_L$, and the mutual information becomes
\small
\begin{equation}
\begin{split}
I(N_{tx}; I_{rx}) =& - \int_{N_{tx}^{min}}^{N_{tx}^{max}} f_{N_{tx}}(x) \Biggl( \log_2\sqrt{2 \pi e \sigma_{I_{rx}}^2} \\
&+ \log_2 f_{N_{tx}}(x) + \log_2\left(\frac{(x \alpha_{ch} k_1 + k_{-1})^2}{N_r M }\right) \Biggr) dx. \label{mutual2}
\end{split}
\end{equation}
\normalsize
\begin{table}[!b]\scriptsize
\centering
\caption{Default Values of Simulation Parameters}
\begin{tabular}{ l | l }
   \hline \hline
   Max number of ligands TN transmits ($N_{tx}^{max}$) & $10^9$  \\ \hline
   Min number of ligands TN transmits ($N_{tx}^{min}$) & $10^8$  \\ \hline
   Transmitter-receiver distance ($d$) & 250 $\mu$m  \\ \hline
%   Diffusion coefficient of ligands ($D_0$) & 10$^{-10}$ m$^2$/s  \\ \hline
   Binding rate ($k_1$) & $2 \times 10^{-18}$ m$^3$/s \\ \hline
   Unbinding rate ($k_{-1}$) & 10 s$^{-1}$  \\ \hline
   Average number of electrons in a ligand ($N_e^-$) & 3  \\ \hline  % protein, DNA
   SiNW radius ($r_{R}$) & 10 nm  \\ \hline
   Concentration of receptors on the surface ($\rho_{SR}$) & $4 \times 10^{16}$ m$^{-2}$ \\ \hline
   Length of a surface receptor ($l_{SR}$) & 2 nm  \\ \hline   % range of aptamer and antibody sizes.
   Temperature ($T$) & $300$K \\ \hline
   Relative permittivity of oxide layer ($\epsilon_{ox}/\epsilon_0$) & $3.9$  \\ \hline
   Relative permittivity of SiNW ($\epsilon_{NW}/\epsilon_0$) & $11.68$  \\ \hline
   Relative permittivity of medium ($\epsilon_R/\epsilon_0$) & $78$  \\ \hline
   Ionic strength of electrolyte medium ($c_{ion}$) & 30 mol/m$^3$  \\ \hline
   Source-drain voltage ($V_{SD}$) & $0.1$ V  \\ \hline
   Source-gate voltage ($V_{SG}$) & $0.4$ V  \\ \hline
   Threshold voltage ($V_{TH}$) & $0$ V  \\ \hline
   Hole density in SiNW ($p$) & $10^{18}$ $cm^{-3}$  \\ \hline
   Tunneling distance ($\lambda$) & $0.05$ nm  \\ \hline
   Thickness of oxide layer ($t_{ox}$) & $2$ nm  \\ \hline
   Oxide trap density ($N_{ot}$) & $10^{16}$ eV$^{-1}$cm$^{-3}$  \\ \hline
   Effective mobility of hole carriers ($\mu_{p}$) & $500$ cm$^2$/Vs  \\ \hline
   Coulomb scattering coefficient ($\alpha_s$) & $1.9 \times 10^{14}$ Vs/C  \\ \hline
\end{tabular}
\label{table:parameters}
\end{table}

To find the capacity, the mutual information should be maximized over all input distributions satisfying $\int_{N_{tx}^{min}}^{N_{tx}^{max}} f_{N_{tx}}(x) dx = 1$. We use the method of Lagrange multipliers defining the Lagrange function as

\small
\begin{equation}
\begin{split}
I_L \equiv I(N_{tx}; I_{rx}) - \lambda \left( \int_{N_{tx}^{min}}^{N_{tx}^{max}} f_{N_{tx}}(x) dx - 1 \right),
\end{split}
\end{equation}
\normalsize
where $\lambda$ is the Lagrange multiplier. The input distribution $f_{N_{tx}}^{*}$ maximizing $I_L$ is then found by functional derivative

\small
\begin{equation}
\begin{split}
 \left. {\frac{\partial I_L}{\partial f_{N_{tx}}(x)}}%
_{\stackunder[1pt]{}{}}%
 \right|_{%
 \stackon[1pt]{$\scriptscriptstyle f_{N_{tx}} = f_{N_{tx}}^{*}$}{}} = 0,
\end{split}
\end{equation}
\normalsize
which gives
%\small
%\begin{equation}
%\begin{split}
%0 = \log_2 \sqrt{2 \pi e \sigma_{I_{rx}}^2} - \frac{\ln f_{N_{tx}}^{*}(x) + 1}{\ln{2}} - \log_2\left(\frac{(x \alpha_{ch} k_1 + k_{-1})^2}{N_r M }\right) - \lambda,
%\end{split}
%\end{equation}
%\normalsize
\small
\begin{equation}
\begin{split}
f_{N_{tx}}^{*}(x) \times \sqrt{2 \pi e \sigma_{I_{rx}}^2} \times \frac{(x \alpha_{ch} k_1 + k_{-1})^2}{N_r M} \times e \times 2^{\lambda} = 1.
\end{split}
\end{equation}
\normalsize

Combining the terms independent of $x$ into a normalization factor $K$, we can write the optimal distribution as follows
\small
\begin{equation}
\begin{split}
f_{N_{tx}}^{*}(x) = \frac{1}{K \sigma_{I_{rx}} (\alpha_{ch} k_1 x + k_{-1})^2 }.
\end{split}
\end{equation}
\normalsize
The normalization factor $K$ can then be obtained as
\small
\begin{equation}
\begin{split}
K =& \int_{N_{tx}^{min}}^{N_{tx}^{max}} \frac{dx}{\sigma_{I_{rx}} (\alpha_{ch} k_1 x + k_{-1})^2} \\
=& \frac{1}{M \sqrt{N_r}} \Biggl[ \sin^{-1}\left( L  \frac{N_{tx}^{max} - k_D/\alpha_{ch}}{N_{tx}^{max} + k_D/\alpha_{ch}}   \right) \\
& - \sin^{-1}\left( L  \frac{N_{tx}^{min} - k_D/\alpha_{ch}}{N_{tx}^{min} - k_D/\alpha_{ch}}  \right) \Biggr],
\end{split}
\end{equation}
\normalsize
where we define
\begin{equation}
L = \sqrt{\frac{g_{FET}^2 \psi_L^2 N_r}{4 \sigma_F^2 + g_{FET}^2 \psi_L^2 N_r}}.
\end{equation}
Substituting $f_{N_{tx}}^{*}(x)$ into \eqref{mutual2}, we obtain the capacity, in bits per use, as follows

\small
\begin{equation}
\begin{split}
C =& \frac{1}{2} \log_2 \frac{N_r}{2 \pi e} + \log_2 \Biggl[ \sin^{-1}\left( L  \frac{N_{tx}^{max} - k_D/\alpha_{ch}}{N_{tx}^{max} + k_D/\alpha_{ch}}  \right) \\
& - \sin^{-1}\left(L  \frac{N_{tx}^{min} - k_D/\alpha_{ch}}{N_{tx}^{min} - k_D/\alpha_{ch}} \right) \Biggr].
\end{split}
\end{equation}
\normalsize

\section{Results}
In this section, we investigate the characteristics of the capacity-achieving input distribution and the effect of main system parameters on the capacity. The default values for the controllable parameters used in the analyses are listed in Table \ref{table:parameters}. The overall setting is the same as that used in \cite{Kuscu2016b}.

%The parameter values are selected assuming that the MC system is exposed to the physiological conditions, and the receptor-ligand pairs correspond to aptamer-natural ligand pairs. The values for parameters related to SiNW and ligand-receptor molecular characteristics, e.g., $N_t$, $t_{ox}$, $R_{mol,L}$, are taken from \cite{Deen2006} \cite{Bang2008} \cite{Rajan2011a} \cite{Hediger2012} \cite{Song2008} \cite{Shorokhov2011} \cite{Landheer2005}; and the range of values for the parameters related to MC system, e.g., $d$, $N_x$, are selected based on the system settings in \cite{Akyildiz2013} \cite{Meng2014} \cite{Pierobon2011}.
\begin{figure}[!t]
\centering
\includegraphics[width=7cm]{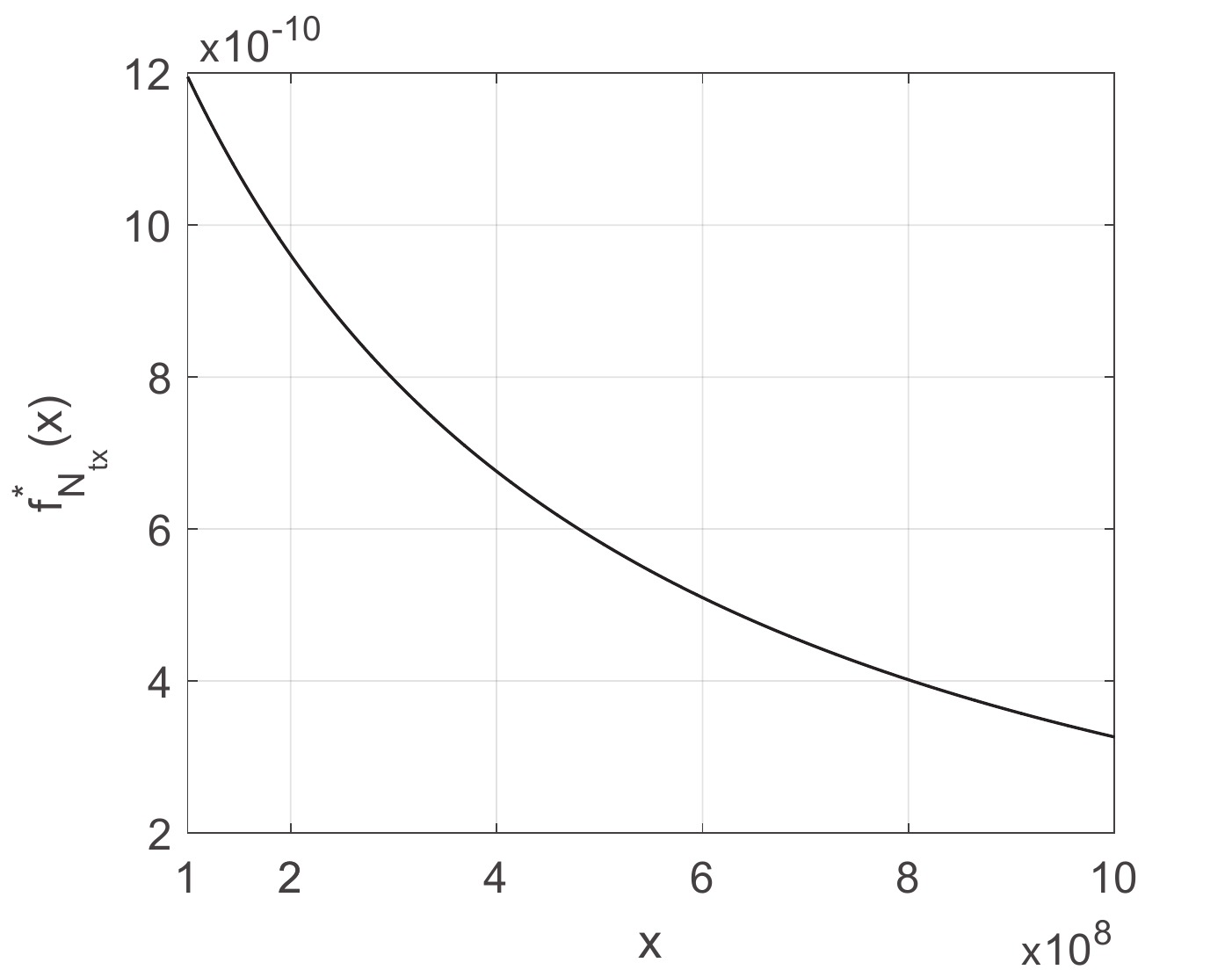}
\caption{Capacity-achieving input distribution with $N_{tx}^{min} = 10^8$ and $N_{tx}^{max} = 10^9$.}
\label{fig:capdist}
\end{figure}

\begin{figure*}[!t]
 \centering
   \subfigure[]{
 \includegraphics[width=4.1cm]{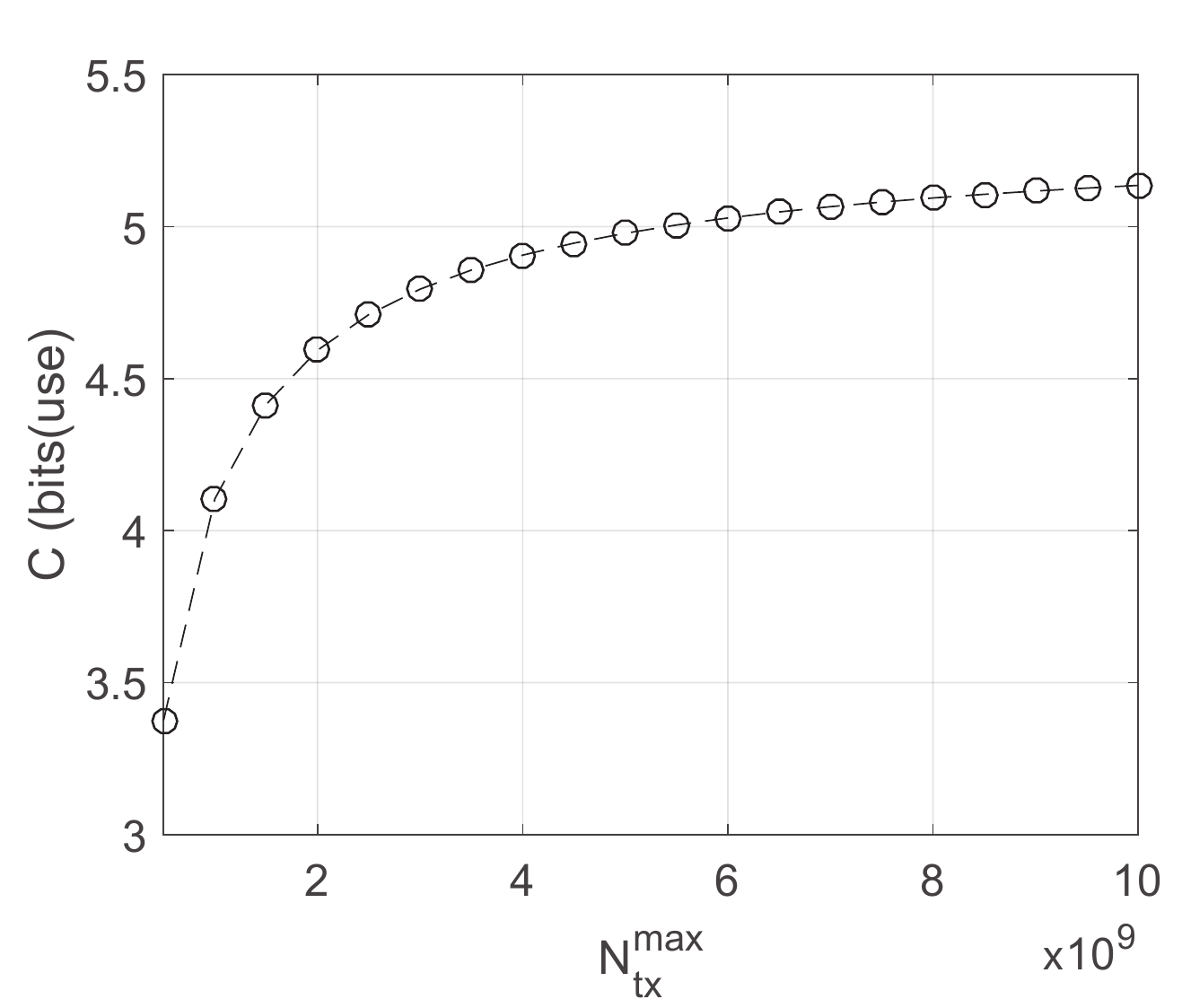}
 \label{fig:capNmax}
 }
  \subfigure[]{
 \includegraphics[width=4.1cm]{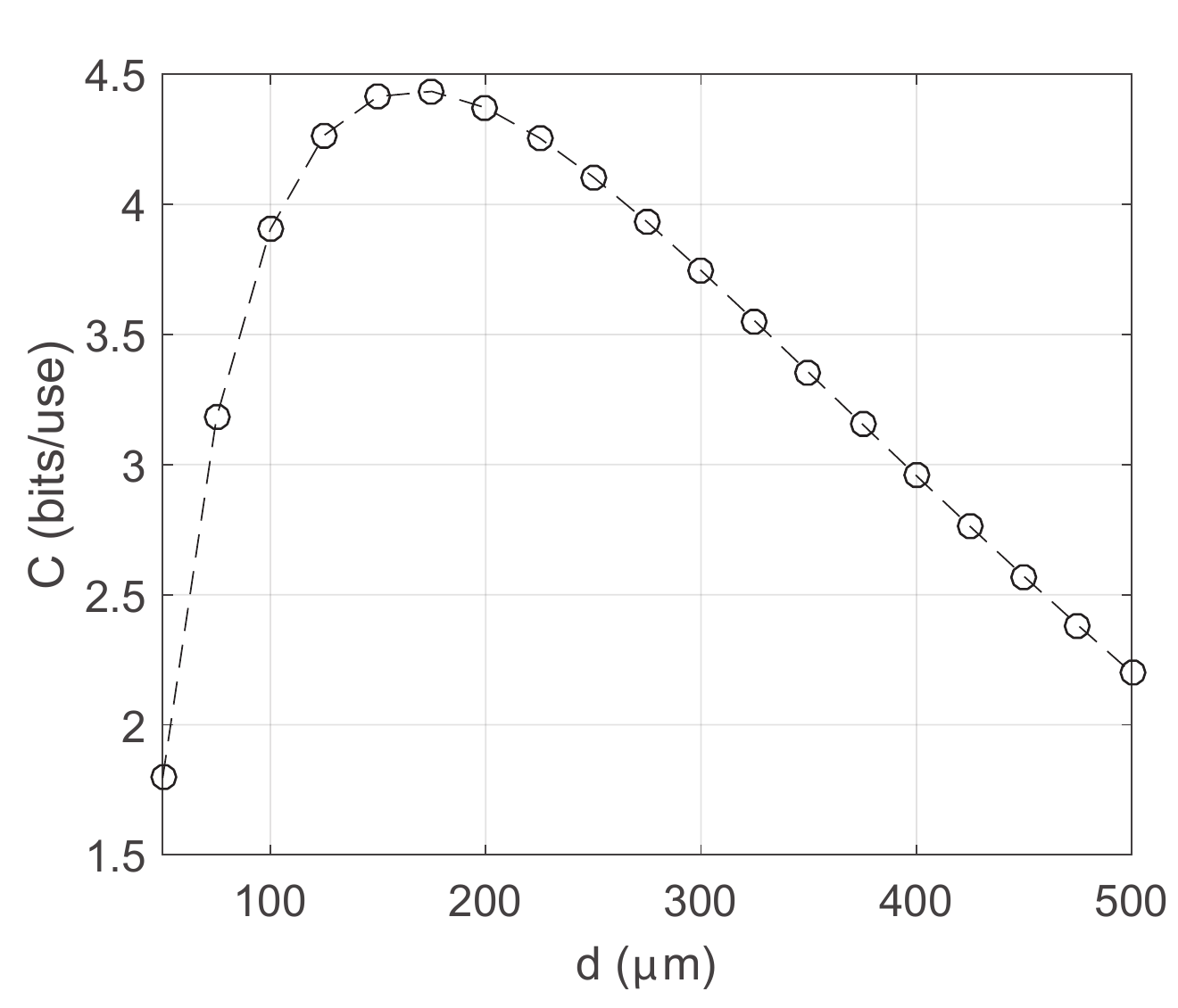}
 \label{fig:capd}
 }
    \subfigure[]{
 \includegraphics[width=4.1cm]{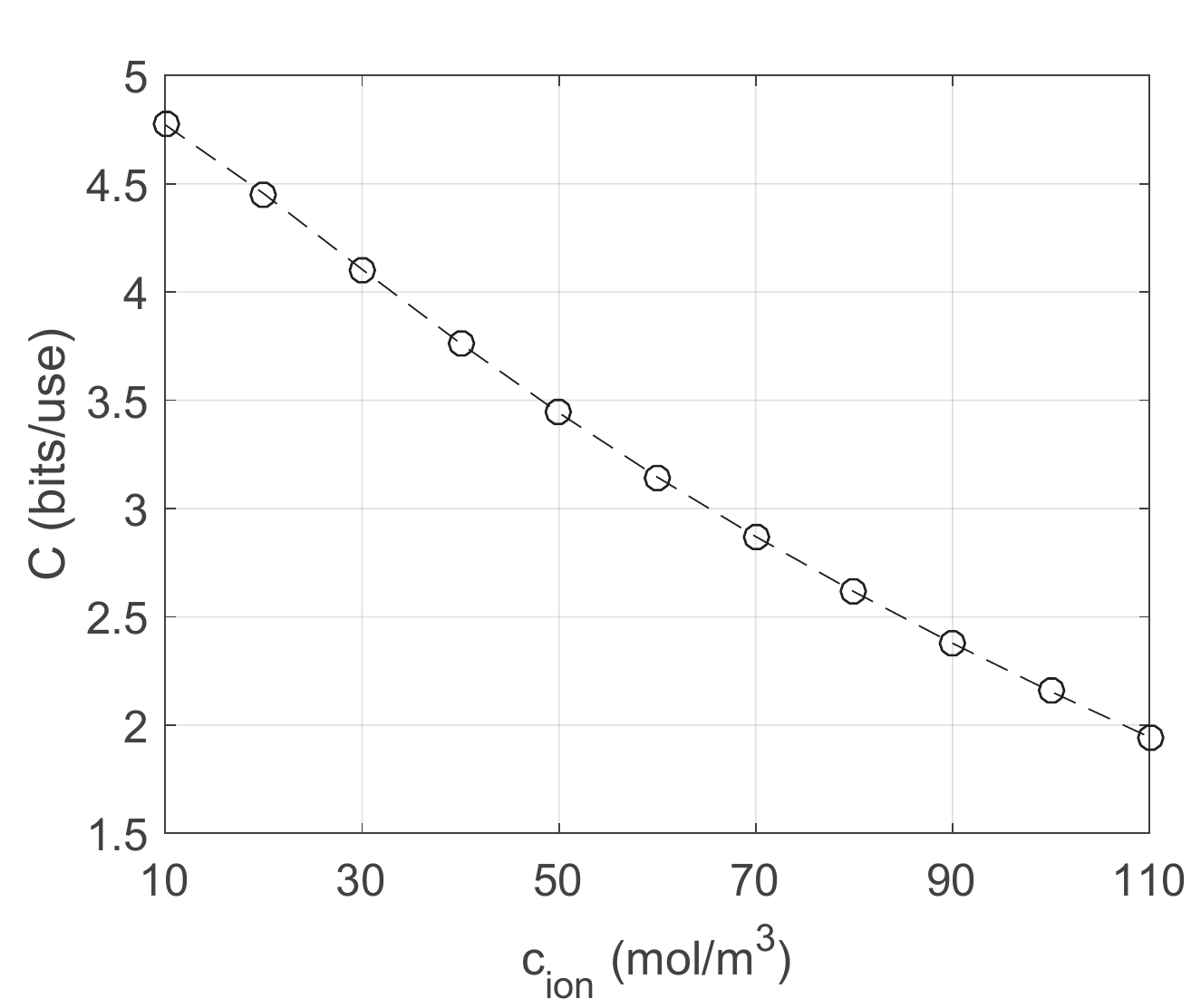}
 \label{fig:capcion}
 }
  \subfigure[]{
 \includegraphics[width=4.1cm]{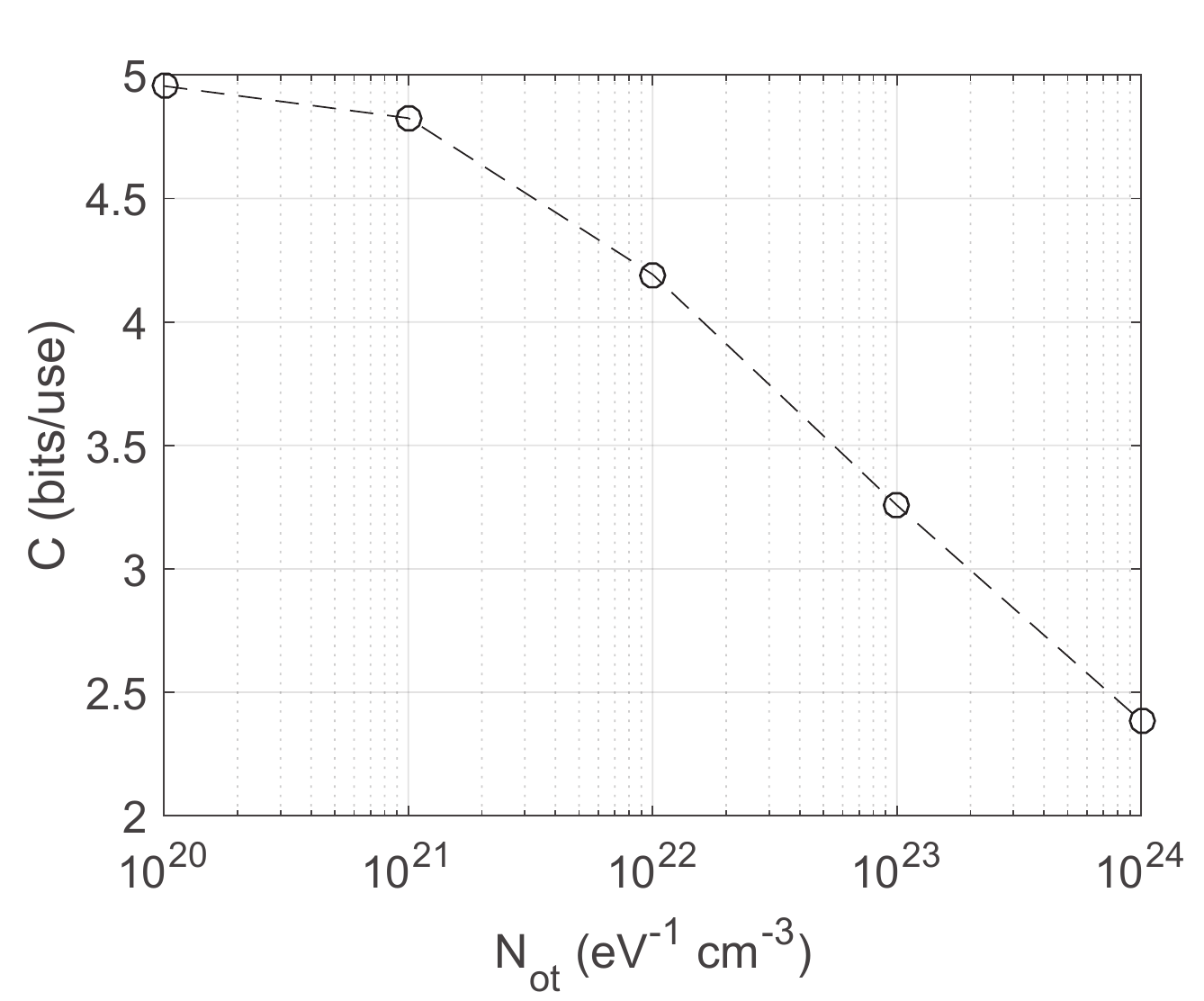}
 \label{fig:capNot}
 }
 \caption{Capacity with (a) varying maximum number $N_{tx}^{max}$ of transmitted molecules, (b) varying TX-RX distance $d$, (c) varying ionic concentration $c_{ion}$, and (d) varying oxide trap density $N_{ot}$.}
 \label{fig:capacity}
 \end{figure*}

The capacity-achieving input distribution $f_{N_{tx}}^{*}$ is demonstrated in Fig. \ref{fig:capdist}. As expected, the distribution is more focused around the minimum number of molecules $N_{tx}^{max}$ TX transmits, which results in a less amount of noise in the receiver's electrical output.

Next, we analyze the relation between the upper limit $N_{tx}^{max}$ of transmitted molecules and the capacity. As is seen in Fig. \ref{fig:capNmax}, the capacity significantly increases with the increasing $N_{tx}^{max}$ until it begins to saturate around 5 bits/use. This is originating from the fact that the receiver has a finite number of molecules; thus, it becomes saturated when $\rho_R \gg K_D$, resulting in a reduced receiver sensitivity. Therefore, transmitting more molecules to increase the ligand concentration at the receiver location is not favorable when the receiver is saturated.

The second analysis is carried out for varying TX-RX distance. As demonstrated in Fig. \ref{fig:capd}, the capacity is maximum at intermediate distances, e.g., $d \approx 150~\mu$m, and begins to decrease when the distance is below or above this range. For a given $N_{tx}^{max}$, as the distance gets smaller, the receiver begins to operate near saturation because $\rho_R$ significantly increases when the transmitter and receiver are close to each other. This is reflected to the output current, and results in a decrease in the sensitivity of the receiver so that it cannot discriminate different levels of ligand concentration corresponding to different symbols. As a result, the capacity decreases. For large distances, the ligand concentration gets significantly attenuated in the channel, thus, the signal-to-noise ratio (SNR) decreases \cite{Kuscu2016b}, resulting in a lower capacity.

The ionic strength of the fluidic medium also substantially affects the capacity, as is seen in Fig. \ref{fig:capcion}. The Debye length $\lambda_D$ decreases with the increasing ionic concentration, implying more effective screening of ligand charges. This in turn reduces the SNR, and thus, the capacity. Physiological conditions generally imply ionic concentrations higher than 100 mol/m$^3$. To compensate the attenuation of capacity, surface receptors with lengths comparable to Debye length should be selected.

Lastly, we analyze the capacity for varying oxide trap density $N_{ot}$, which is proportional to the impurity of the SiNW. Trap density affects the carrier mobility, and increases the $1/f$ noise. The negative effect of increasing trap density on the capacity is evident from Fig. \ref{fig:capNot}.

\section{Conclusion}

In this study, we developed an information theoretical model for a memoryless MC system equipped with a SiNW FET-based MC receiver, and derived analytical closed-form expressions for the capacity and capacity-achieving input distribution. The provided expressions enable the analysis of the effect of nanobioelectronic receiver parameters on the MC system capacity, and reveal the optimization pathways that can be targeted to improve reliability of the overall communication system.

%In this paper, as the first step towards implementing a human-made MC system with nanobioelectronic devices, we have developed a communication theoretical model for SiNW FET-based MC receivers integrating all the underlying processes in MC and bioFET operation. We have derived closed-form expressions for fundamental performance metrics, such as SNR and SEP, to provide an analysis and optimization framework for MC with nanobioelectronic receivers. The results of numerical analyses have pointed out several optimization pathways that need to be taken to improve the detection performance of the receiver. The developed model can be extended to incorporate the transient dynamics of MC and bioFETs for enabling analysis also in the frequency domain. Open issues include the design of optimal constellations and optimal receiver detection schemes for MC systems equipped with bioFET receivers. Further research on devising nanobioelectronic MC receivers could enable the implementation of all the theoretical protocols and algorithms designed for reliable and efficient MC and the development of seamless interfaces between MC nanonetworks and macroscale networks towards realizing IoNT.
%\section*{Acknowledgment}
%This work is a part of the Project MINERVA supported by the European Research Council (ERC) under grant ERC-2013-CoG \#616922.


\begin{thebibliography}{1}
\bibitem{Akan2016}
O. B. Akan, H. Ramezani, T. Khan, N. A. Abbasi and M. Kuscu,
``Fundamentals of molecular information and communication science,"
\emph{Proc. IEEE,} to be published, 2016.

\bibitem{Kuscu2015}
M. Kuscu and O. B. Akan,
``The Internet of molecular things based on FRET,"
\emph{IEEE Internet Things J.,} vol. 3, no. 1, pp. 4-17, 2016.

\bibitem{Akyildiz2013}
M. Pierobon and I. F. Akyildiz,
``Capacity of a diffusion-based molecular communication system with channel memory and molecular noise,"
\emph{IEEE Trans. Inf. Theory,} vol. 59, no. 2, pp. 942-954, 2013.

\bibitem{Nakano2012}
T. Nakano et al.,
``Molecular communication and networking: opportunities and challenges,"
\emph{IEEE Trans. Nanobiosci.,} vol. 11, no. 2, pp. 135-148, 2012.

\bibitem{Unluturk2015}
B. D. Unluturk et al.,
``Genetically engineered bacteria-based biotransceivers for molecular communication,"
\emph{IEEE Trans. Comm.,} vol. 63, no. 4, pp. 1271-1281, 2015.

\bibitem{Kuscu2016a}
M. Kuscu and O. B. Akan,
``On the physical design of molecular communication receiver based on nanoscale biosensors,"
\emph{IEEE Sensors J.,} vol. 16, no. 8, pp. 2228-2243, 2016.

\bibitem{Kuscu2016b}
M. Kuscu and O. B. Akan,
``Modeling and analysis of SiNW FET-based molecular communication receiver,"
\emph{IEEE Trans. Comm.,} to be published, 2016.

\bibitem{Marzen2013}
S. Marzen, H. G. Garcia and R. Phillips,
``Statistical mechanics of Monod-Wyman-Changeux (MWC) models,"
\emph{J. Mol. Biol.,} vol. 425, no. 9, pp. 1433-1460, 2013.

\bibitem{llatser2011}
I. Llatser et al.,
``Diffusion-based channel characterization in molecular nanonetworks,"
in \emph{Proc. IEEE INFOCOM,} Shangai, China, April 2011.

\bibitem{Berezhkovskii2013}
A. M. Berezhkovskii and A. Szabo,
``Effect of ligand diffusion on occupancy fluctuations of cell-surface receptors,"
\emph{J. Chem. Phys,} vol. 139, pp. 121910, 2013.

\bibitem{Rajan2013}
N. K. Rajan et al.,
``Performance limitations for nanowire/nanoribbon biosensors,"
\emph{Wiley Interdiscip. Rev. Nanomed. Nanobiotechnol.,} vol. 5, no. 6, pp. 629-645, 2013.

\bibitem{Gao2010}
X. P. A. Gao, G. Zheng and C. M. Lieber,
``Subthreshold regime has the optimal sensitivity for nanowire FET biosensors,"
\emph{Nano Lett.,} vol. 10, pp. 547-552, 2010.

\bibitem{Shoorideh2014}
K. Shoorideh and C. O. Chui,
``On the origin of enhanced sensitivity in nanoscale FET-based biosensors,"
\emph{PNAS,} vol. 111, no. 4, pp. 5111-5116, 2014.

\bibitem{Rajan2010}
N. Rajan et al.,
``Temperature dependence of $1/f$ noise mechanisms in silicon nanowire biochemical field effect transistors,"
\emph{Appl. Phys. Lett.,} vol. 97, pp. 243501, 2010.

\bibitem{Niemann2013}
M. Niemann, H. Kantz and E. Barkai,
``Fluctuations of 1/f noise and the low-frequency cutoff paradox,"
\emph{Phys. Rev. Lett.,} vol. 110, no. 140603, 2013.

\bibitem{Hooge1969}
F. N. Hooge, A. M. H. Hoppenbrouwers,
``Amplitude distribution of 1/f noise,"
\emph{Physica,} vol. 42, no. 3, pp. 331-339, 1969.



\end{thebibliography}
\end{document}